\documentclass{jpsj2}
\usepackage{graphicx}

\title{Anisotropy of Resonant Inelastic X-Ray Scattering at the $K$ Edge of Si:\\Theoretical Analysis}

\author{Yunori {\sc Nisikawa}$^{1}$\footnote{E-mail address:nisikawa@spring8.or.jp}, 
Manabu {\sc Usuda}$^{1}$ and Jun-ichi {\sc Igarashi}$^{2}$}

\inst{$^{1}$Synchrotron Radiation Research Center, Japan Atomic Energy
Research Institute, Mikazuki, Sayo, Hyogo 679-5148 \\

$^{2}$Faculty of Science, Ibaraki University, Mito,
Ibaraki 310-8512}

\recdate
{
\today
}

\abst{We investigate theoretically the resonant inelastic x-ray 
scattering (RIXS) at the $K$ edge of Si on the basis of an ab initio
calculation. We calculate the RIXS spectra
with systematically varying transfered-momenta, 
incident-photon energy and incident-photon polarization.
We confirm the anisotropy of the experimental spectra
by Y. Ma {\it et al}. (Phys. Rev. Lett. 74, 478 (1995)),
providing a quantitative explanation of the spectra.
}

\kword{resonant inelastic X-ray scattering, $K$ edge of Si}

\begin{document}
\maketitle
\section{Introduction} 
Inelastic x-ray scattering is a promising method
to study electronic structures in matters.
It is advantageous to use a resonant enhancement
by tuning photon energy near absorption edge.
Recently it has been revealed that the resonant inelastic
x-ray scattering (RIXS) is a powerful tool for elucidating
the electronic properties of solids. 
RIXS measurement has been
applied to search for charge excitations 
in several cuprates\cite{HASAN,KIM} and manganites\cite{INAMI}, 
with tuning incident photon energy to the absorption K edge.
Since corresponding photon energies are in the hard-X-ray region, 
the momentum transfer can not be neglected.
Analyzing the transfered-momentum dependence of the RIXS spectra,
we can clarify the characteristics of charge excitations
in strongly-correlated electron systems.

In semiconductors such as Si and Ge, the interpretation of the RIXS spectra
may be simpler because of weak electron correlations.
The final state of RIXS consists of one electron in the conduction band and
one hole in the valence band. 
With neglecting the interaction between the electron and the hole,
one can draw useful information of single-particle spectra
from the transferred-momentum dependence.
Actually Ma has tried to determine valence band structures of Si 
by systematically varying scattering vectors in the RIXS experiment.
\cite{rf:MaTheory}

We have already carried out a RIXS experiment at the $K$ edge of Ge 
with the same aim.
Although we systematically varied transfered-momenta,
we obtained merely a broad inelastic peak as a function of photon energy,
whose shape was nearly independent of transfered-momenta.
Performing the band structure calculation,
we obtain the spectral shape in agreement with the experiment,
and clarified the origin of the spectral shape\cite{rf:Netal}.
This analysis led us to realize that
the decisive factor whether the spectra are depending on the transferred
momenta or not is the core-level width.
If the core-hole level in the intermediate state is sharp enough, 
we can definitely select the momentum of the excited electron 
in the conduction band 
by sharply tuning the incident-photon energy, and thereby 
we can also specify the momentum of the hole in the valence band 
by setting the scattering geometry. In this situation,
sharp peaks are expected to be observed as a function of scattered
photon energy.
If the core-level width is not small,
many channels of exciting an electron with different momenta 
in the conduction band are expected to be opened. Thereby, sharp peaks
are blurred up by overlapping each contribution.
In an extreme situation, the spectra become close to a superposition 
of the density of states (DOS) projected onto $p$-symmetric states.
Unfortunately, for Ge, the $1s$ core-level width is as large as $\sim 2$eV, 
an order of the width of conduction band, so that
the RIXS spectra did not show clear momentum dependence.

For Si, since the $K$ edge energy is around 1840 eV,  
the transfered-momentum cannot be neglected in the RIXS experiment.
The $1s$ core-level width is estimated about 0.6 eV, which is 
smaller than the width of conduction band of Si.
%
These facts imply a possibility of observing the transfered-momentum 
dependence at the $K$ edge in the RIXS spectra.
Actually, Ma {\it et. al} observed such transfered-momentum dependence
\cite{rf:MaExp}.
In this paper, we analyze Ma's experimental RIXS spectra
with the help of {\em ab initio} band structure calculation,
and elucidate the relation between the spectral shape and the underlying
electronic structure.
%

This paper is organized as follows. In \S \ref{forma},
the formalism for calculating the spectra is described.
In \S \ref{caldis}, the calculated results and discussion are presented. 
Section \ref{conc} is devoted to concluding remarks.
\section{Formulation}\label{forma}

RIXS is described by a second-order process
that the incident photon with energy $\hbar\omega_1$, 
momentum $\hbar{\bf q}_1$ and polarization ${\bf e}_1$ is virtually
absorbed by exciting the core electron to the conduction band and then
a photon with energy $\hbar\omega_2$,
momentum $\hbar{\bf q}_2$ and polarization ${\bf e}_2$
is emitted by filling the core-hole state with a valence electron.
The scattering geometry is shown in Fig.~\ref{fig:arange}.
The normal axis ${\bf n}$ and theta axis ${\bf t}$ in this figure
are respectively defined as follows:
${\bf n}\propto{\bf q}_{2}/|{\bf q}_{2}|-{\bf q}_{1}/|{\bf q}_{1}|$,
 ${\bf t}\propto {\bf q}_{1}\times {\bf q}_{2}$.
 The setup for scattering is uniquely determined
by choosing the normal and theta axis from crystal axes of Si.

To obtain the double differential scattering cross-section, 
we use the generalized Fermi's golden rule where the interaction between 
photon and electrons is treated by second order perturbation theory.
An independent-particle approximation seems to be appropriate,
since electron correlations are expected to be weak~\cite{rf:MaTheory}.
In such a situation, the double differential scattering cross-section 
may be expressed as
\begin{equation}\label{dsS}
\frac{d^{2}\sigma}{d\omega_{2} d\Omega_{2}}\propto
\sum_{({\bf k},e),({\bf k}^{\prime},h)}
\frac
{\left|\sum_{a}\exp(i{\bf \Omega}\cdot{\bf R}_{a})\overline{t_{a}({\bf k}^{\prime},h|{\bf e}_{2})}
t_{a}({\bf k},e|{\bf e}_{1})\right|^{2}}
{(\epsilon_{e}({\bf k})-\epsilon_{c}-\hbar\omega_{1})^{2}+\Gamma^{2}/4}
\delta^{{\bf G}}_{{\bf \Omega},{\bf k}-{\bf k}^{\prime}}
\delta(\epsilon_{e}({\bf k})-\epsilon_{h}({\bf k}^{\prime})-\hbar\omega),
\end{equation}
where $\hbar\omega=\hbar\omega_{1}-\hbar\omega_{2}$, and $\hbar{\bf
 \Omega}=\hbar{\bf  q}_{1}-\hbar{\bf q}_{2}$ are energy and momentum of the
 final state.
$\epsilon_{e}({\bf k})$, $\epsilon_{h}({\bf k}^{\prime})$ and $\epsilon_{c}$
are the energy of the excited electron with crystal momentum ${\bf k}$
in the conduction band $e$, that of the hole with crystal momentum
${\bf k}^{\prime}$ in the valence band $h$, and that of the core state,
respectively.
Overlined quantity indicates the complex conjugate.
The $\Gamma$ is the $1s$ core-level width of Si.
Quantity $t_{a}({\bf k}^{\prime},h|{\bf e}_{2})\equiv
\int d{\bf r}\overline{\psi_{{\bf k}^{\prime},h}({\bf r})}
{\bf e}_{2}\cdot\hat{{\bf p}}\phi_{a}^{1s}({\bf r}-{\bf R}_{a})$ 
describes the transition from the valence band to the $1s$ core state,
where ${\bf R}_{a}$, $\psi_{{\bf k}^{\prime},h}$
and $\phi_{a}^{1s}$ are the position
vector of atom $a$ in unit cell,
Bloch-wave function of an electron in the valence band $h$ with
crystal momentum ${\bf k}^{\prime}$,
and 1$s$-atomic orbital, respectively.
The crystal momentum conservation for the whole process is contained
in the factor of Kronecker $\delta$,
\begin{equation}
\delta^{\bf G}_{{\bf \Omega},{\bf k}-{\bf k}^{\prime}}
\equiv  \left\{
\begin{array}{@{\,}ll}
0 & \mbox{: ${\bf \Omega}-({\bf k}-{\bf k}^{\prime})\notin {\bf G}$ }\\
1 & \mbox{: ${\bf \Omega}-({\bf k}-{\bf k}^{\prime})\in {\bf G}$ }
\end{array}
\right.,
\end{equation}
where ${\bf G}$ is the set of reciprocal lattice vectors. 

\section{Calculated Results and Discussion}\label{caldis}

\subsection{Band Structure Calculation of Si}
We perform a band structure calculation using 
the full-potential linearized augmented-plane-wave (FLAPW) method 
within the local-density approximation (LDA).
The local exchange-correlation functional of Vosko, Wilk and Nusair
 is employed~\cite{VWN80}. 
The angular momentum in the spherical-wave expansion is truncated at
 $l_{\rm max}=6$ and $7$ for the potential and wave function, respectively.
 The energy cutoff of the plane wave is 12 Ry for the wave function.
Figure \ref{fig:band} shows the energy vs. momentum 
relation thus evaluated.  
The energy band is labeled by attached numbers for later use.



%
%

\subsection{RIXS spectra}\label{aniso}
\subsubsection{Anisotropy of RIXS at $K$ edge of Si}
We first analyze the transfered-momentum dependence of RIXS 
spectra for the incident photon energy $\hbar\omega_{1}=\epsilon_{\rm
min}-\epsilon_{c}$, where $\epsilon_{\rm min}$ is the minimum energy of
conduction-bands.
When $\Gamma$ is very small, the denominator in Eq.~\ref{dsS} forces 
the excited electron stay just at the bottom 
of the conduction band. There are six such points, which are denoted as
${\bf k}_{i}$ $(i=1, \cdots, 6)$ in Fig.~\ref{fig:BZ}. They 
belong to band $5$ (see Fig.\ref{fig:band}).
We can specify the momentum of hole ${\bf q}_i\equiv {\bf k}_i-{\bf\Omega}$
in the valence band by specifying the scattering vector. 
In such a situation, the double differential scattering cross-section
can be rewritten as 
\begin{eqnarray}\label{dsSsmallG}
\frac{d^{2}\sigma}{d\omega_{2} d\Omega_{2}}&\propto&
\sum_{i, h}
A(i,h|{\bf \Omega},{\bf e}_{1})
\delta(\epsilon_{\rm min}-\epsilon_{h}({\bf q}_{i})-\hbar\omega),\\
A(i,h|{\bf \Omega},{\bf e}_{1})&\equiv&\sum_{{\bf e}_{2}}
\left|\sum_{a}e^{i{\bf \Omega}\cdot{\bf R}_{a}}\overline{t_{a}
({\bf q}_{i},h|{\bf e}_{2})}
t_{a}({\bf k}_{i},5|{\bf e}_{1})\right|^{2}.\label{amp}
\end{eqnarray}
The sharp peaks are expected at 
$\hbar\omega=\hbar\omega(i,h)\equiv\epsilon_{\rm min}
-\epsilon_{h}({\bf q}_{i})$
in the spectra, where $h$ $(=1,\cdots, 4)$ indicates the valence band index.
When the scattering vector is along a symmetry line, the number of peaks
is reduced because of degeneracy of peaks and no intensities to some peaks.
%
For ${\bf n}=(100)$, for example, the momenta of the holes
are shown in Fig.\ref{fig:BZ},
and the corresponding valence band energies are the same due to symmetry,
that is, $\epsilon_{h}({\bf q}_{2})=\epsilon_{h}({\bf q}_{3})=
\epsilon_{h}({\bf q}_{4})=\epsilon_{h}({\bf q}_{5})$ with $h=1, \cdots 4$,
except for ${\bf q}_1$ and ${\bf q}_6$. 
Therefore, several excitation energies coincide with each other,
and thereby different peak positions are
$\hbar\omega=\hbar\omega(1,h), \hbar\omega(2,h),\hbar\omega(6,h)$
with $h=1, \cdots, 4$.
In addition, the spectral weights 
$A(i,h|{\bf \Omega},{\bf e}_{1}=\sigma)$ for momentum 
${\bf k}_1$ and ${\bf k}_6$
are found to be almost zero.
Thus only four peaks appear at $\hbar\omega=\hbar\omega(2,h) (h=1,\cdots, 4)$,
as shown in Fig. ~\ref{fig:G0vsG0.6} (c).
With slight increase of the $\Gamma$ value, the excited electron can stay 
at several states near the bottom of the conduction band,
and thereby the valence hole can stay at several states 
near ${\bf q}_i(i=2, \cdots, 5)$.
Therefore the spectral peaks are obscured.
We can make similar analysis for ${\bf n}= (111)$, $(110)$.
Such spectral changes are demonstrated in Fig.\ref{fig:G0vsG0.6}
for ${\bf n}= (111)$, $(110)$ and $(100)$, with
$\Gamma$ changed from $0.02$ eV  to $0.6$ eV.

When $\Gamma$ is large, the momentum of the excited electron can take whole
values in the first Brillouin zone, and so does the momentum of the hole 
in the valence band. The spectral peaks are smeared out, and thereby 
becoming close to the DOS projected onto $p$ symmetric states. 
Thus we have the spectra almost independent of scattering vector. 
This behavior is demonstrated in Fig.\ref{fig:Gc}, where $\Gamma=1.6$ eV.

For the actual Si case, $\Gamma$ is close to $0.6$ eV,
and the spectral shape is expected to show the transferred-momentum
dependence.
Figure \ref{fig:1840.8} shows the RIXS spectra with $\Gamma=0.6$ eV
for $\hbar\omega_{1}=1840.8$ eV, 
in comparison with experimental results by Ma {\it et. al}\cite{rf:MaExp}.
In the experimental setup of Ma {\it et. al}\cite{rf:MaExp},
the normal axis ${\bf n}$ is (111), (110), and (100), 
but the theta axis ${\bf t}$ and incident-photon polarization 
are not clearly indicated in their paper\cite{rf:MaExp}.
Assuming the $\sigma$ polarization for the incident photon 
and the theta axis ${\bf t}$ as given in the figure,
we are successful in reproducing the experimental shape 
such as three peak structure for ${\bf n}=(111)$
and the different spectral shape for ${\bf n}=(110)$ and $(100)$.
Corresponding to the above discussion, we can identify where the peak
structures come from by decomposing the spectra into each contribution 
from the hole band and the electron band.
Figure \ref{fig:decomp} shows such decomposition, where attached numbers
indicates the band index given in Fig.~\ref{fig:band}.

\subsubsection{Incident-photon energy dependence of RIXS spectra}
The spectra strongly depend on the incident-photon energy.
Figure \ref{fig:enedep111} shows the spectra with varying incident-photon 
energy from 1840.0 eV to 1841.2 eV.
The scattering vector is fixed to be proportional to ${\bf n}=(111)$.
The characteristic three peaks A, B and C are visible only in the narrow 
energy region between 1840.2 eV and 1840.8 eV.
For $\hbar\omega_{1}=1842.5$ eV, the anisotropy almost disappears,
as shown in Fig.~\ref{fig:enedephigh}.
The result is consistent with experimental results 
of Ma {\it et. al}\cite{rf:MaExp}.
 
\subsubsection{Incident-photon polarization dependence} 
Figure \ref{fig:poldep} shows the RIXS spectra calculated
for $\sigma$ and $\pi$ incident polarizations. 
It is found that 
the difference in RIXS spectra in higher $\hbar\omega_{2}$ 
for ${\bf n}=(100)$ is large, 
while not for ${\bf n}=(111)$ and $(110)$.
This tendency is more clearly seen in the limit situation $\Gamma=0.02$ eV 
where $\hbar\omega_{1}=\epsilon_{\rm min}-\epsilon_{c}$, 
as shown in Fig.\ref{fig:kstar}.
The positions of peaks expected to be made by 
each electron-hole pairs are also presented in Fig.\ref{fig:kstar}.
To make the discussion clear, we consider 
the limit situation mentioned above. 
Then, the $1s$ electron is excited to the states with momentum ${\bf k}_i$
$(i=1,\cdots,6)$. Corresponding wave functions projected onto
p-symmetric states centering on ${\bf R}_{a}(a=1, 2)$ are given by
$P_{l=1}\psi_{{\bf k}_{1},5}\propto p_{x}$,
$P_{l=1}\psi_{{\bf k}_{2},5}\propto p_{y}$,
$P_{l=1}\psi_{{\bf k}_{3},5}\propto p_{y}$,
$P_{l=1}\psi_{{\bf k}_{4},5}\propto p_{z}$,
$P_{l=1}\psi_{{\bf k}_{5},5}\propto p_{z}$,
$P_{l=1}\psi_{{\bf k}_{6},5}\propto p_{x}$,
as shown in the left side of Fig.\ref{fig:kstar}.
In the case of ${\bf n}=(100)$, ${\bf t}=(011)$, 
the incident-photon polarization vector 
${\bf e}_1$ is proportional to $(0,1,1)$ in the $\sigma$ polarization
and $(\sqrt{2},-1,1)$ in the $\pi$ polarization.
The $x$-component of ${\bf e}_1$ is not zero only in the case of 
$\pi$ polarization.
Then the $t_{a}({\bf k}_{i},5|{\bf e}_{1})$ in Eq.\ref{amp} 
shows a large change 
at ${\bf k}_{1}$ and ${\bf k}_{6}$, with changing polarization;
$|t_{a}({\bf k}_{1,6},5|\pi)|>>|t_{a}({\bf k}_{1,6},5|\sigma)|\simeq 0$, 
$|t_{a}({\bf k}_{2,\cdots,5},5|\pi)|\simeq|t_{a}({\bf k}_{2,\cdots, 5},5|\sigma)|$.
Therefore, we can expect that the electron-hole pairs 
with momenta ${\bf k}_1$, ${\bf q}_1$ 
and ${\bf k}_6$, ${\bf q}_6$ contribute to RIXS spectrum in the case of 
the $\pi$ polarization. 
Note that the positions of the peaks made by 
the electron-hole pairs with momenta ${\bf k}_1$, ${\bf q}_1$ 
and ${\bf k}_6$, ${\bf q}_6$ in the $\pi$ polarization 
are quite different from 
these of the peaks made by other electron-hole pairs 
with momenta ${\bf k}_i$, ${\bf q}_i (i=2,\cdots, 5)$.
From Fig.~\ref{fig:kstar}(c), we find that 
electron-hole pairs with momenta ${\bf k}_1$, ${\bf q}_1$ for $h=2,3,4$, 
and ${\bf k}_6$, ${\bf q}_6$ for $h=3, 4$ significantly contribute to forming
the peak around $\hbar\omega_2=1839$ eV in the $\pi$ polarization. 
Such a contribution gives rise to the appreciable intensity 
around $\hbar\omega_2=1839$ eV in the realistic RIXS spectrum for 
the $\pi$ polarization, as 
shown in Fig.~\ref{fig:poldep}.
The same discussion mentioned above can be applied in the case of 
${\bf n}=(111)$, ${\bf t}=$(1-10) and ${\bf n}=(110)$, ${\bf t}=$(1-10).
In the case of 
${\bf n}=(111)$, ${\bf t}=$(1-10) and ${\bf n}=(110)$, ${\bf t}=$(1-10),
the incident-photon polarization vector ${\bf e}_1$ is 
proportional to $(1,-1,0)$ in the $\sigma$ polarization
and $(1,1,\alpha), (\alpha\neq 0)$ in the $\pi$ polarization.
The $z$-component of ${\bf e}_1$ is not zero only in the case of 
$\pi$ polarization.
Therefore, the $t_{a}({\bf k}_{i},5|{\bf e}_{1})$  
shows a large change 
at ${\bf k}_{4}$ and ${\bf k}_{5}$, with changing the polarization;
$|t_{a}({\bf k}_{4,5},5|\pi)|>>|t_{a}({\bf k}_{4,5},5|\sigma)|\simeq 0$, 
$|t_{a}({\bf k}_{1,2,3,6},5|\pi)|\simeq|t_{a}({\bf k}_{1,2,3,6},5|\sigma)|$.
Therefore, we can expect that the electron-hole pairs 
with momenta ${\bf k}_4$, ${\bf q}_4$ 
and ${\bf k}_5$, ${\bf q}_5$ contribute to RIXS spectrum in the case of 
the $\pi$ polarization. 
From Fig. ~\ref{fig:kstar}(a),(b), we can see the following facts.
In the case of ${\bf n}=(111)$, ${\bf t}=$(1-10),
the positions of the peaks made by the 
electron-hole pairs with momenta ${\bf k}_4$, ${\bf q}_4$  
and ${\bf k}_5$, ${\bf q}_5$ in the $\pi$ polarization
are same as these of the peaks made by other 
electron-hole pairs with momenta 
${\bf k}_i$, ${\bf q}_i (i=1, 2, 3, 6)$.
In the case of ${\bf n}=(110)$, ${\bf t}=$(1-10), 
the position of the peak made by the 
electron-hole pairs with momenta ${\bf k}_4$, ${\bf q}_4$  
and ${\bf k}_5$, ${\bf q}_5$ for $h=2$ in the $\pi$ polarization
is same as that of the main peak in the $\sigma$ polarization. 
On the other hand, the positions of the peaks made by the 
electron-hole pairs with momenta ${\bf k}_4$, ${\bf q}_4$  
and ${\bf k}_5$, ${\bf q}_5$ for $h=3,4$ in the $\pi$ polarization 
are different from that of the peaks made by other electron-hole pairs 
with momenta ${\bf k}_i$, ${\bf q}_i (i=1, 2, 3, 6)$. 
But the new peaks made by 
electron-hole pairs with momenta  ${\bf k}_4$, ${\bf q}_4$  
and ${\bf k}_5$, ${\bf q}_5$ for $h=3,4$ are not so significant.
Summarizing, in the case of ${\bf n}=(111)$, ${\bf t}=$(1-10) and 
${\bf n}=(110)$, ${\bf t}=$(1-10), significant peaks at new positions
are not expected by varying incident-photon polarization.
Therefore, the realistic RIXS spectra 
for ${\bf n}=(111)$, ${\bf t}=$(1-10) and 
${\bf n}=(110)$, ${\bf t}=$(1-10) are not sensitive to 
the incident-photon polarization, as shown 
in Fig.~\ref{fig:poldep}.

\section{Concluding Remarks}\label{conc}
We have calculated the RIXS spectra at the $K$ edge of Si 
on the basis of the band structure calculation.
We have successfully reproduced the anisotropy of the RIXS spectra
in the experiment by Ma {\it et. al}\cite{rf:MaExp}.
We have analyzed the spectra with systematically varying
transfered-momenta, incident-photon energy and
incident-photon polarization, thus having provided a quantitative explanation
of the spectra.
With the help of the band calculation, 
one can draw information on the valence band structure from the RIXS spectra,
although it is indirect.
This information seems helpful with considering that the RIXS is 
bulk-sensitive in contrast to photoemission with a lower energy photon 
which is surface-sensitive.

Finally, we comment the following general tendencies.
To have an anisotropy in RIXS at $K$ edge of specific element in
semiconductor or band-insulator, the sizable momentum transfer 
and core-level width smaller than conduction-bandwidth are required.
The photon energy is required to obtain sizable momentum
transfer but corresponding core-level widths are usually large. 
Therefore, it is difficult to satisfy the requirement of 
the sizable momentum transfer and small core-level width.

\section*{Acknowledgment}
Authors greatly thank to 
N. Hamada for allowing us to use his FLAPW code,
and to T. Iwazumi and A. Agui for valuable discussion.
This work was supported in part by a Grant-in-Aid for Scientific Research
from the Ministry of Educations, Science, Sports, and Culture.

\clearpage
\setcounter{figure}{0}
\begin{figure}
\includegraphics[width=15cm,height=13cm]{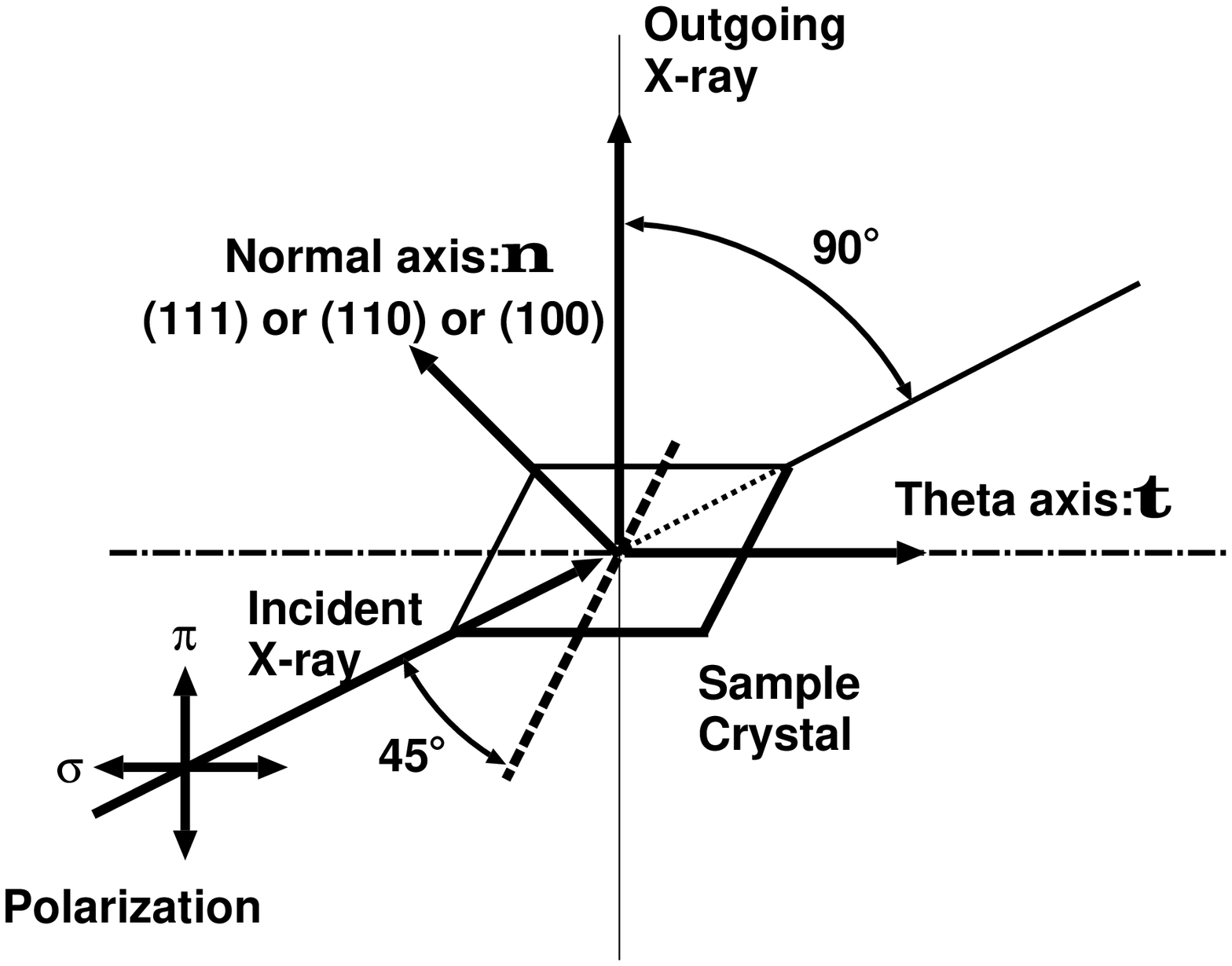}
\caption{Geometrical relations
 between ${\bf q}_{1}, {\bf q}_{2}$ and crystal axis of Si. }
\label{fig:arange}
\end{figure}

\begin{figure}
\includegraphics[width=15cm,height=13cm]{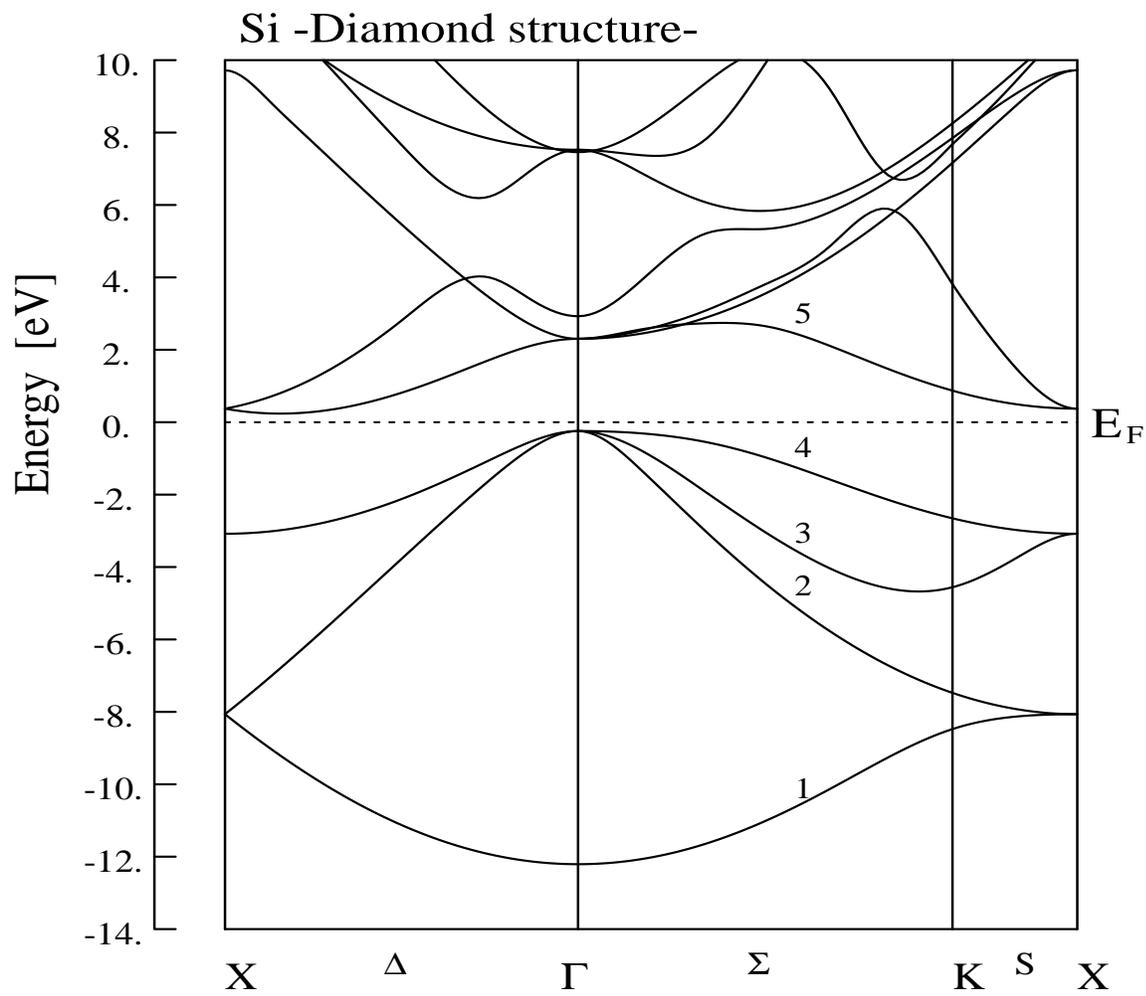}
\caption{Energy vs. momentum relation of Si calculated by the FLAPW method
in the LDA scheme.}
\label{fig:band}
\end{figure}


\begin{figure}
\includegraphics[width=15cm,height=11cm]{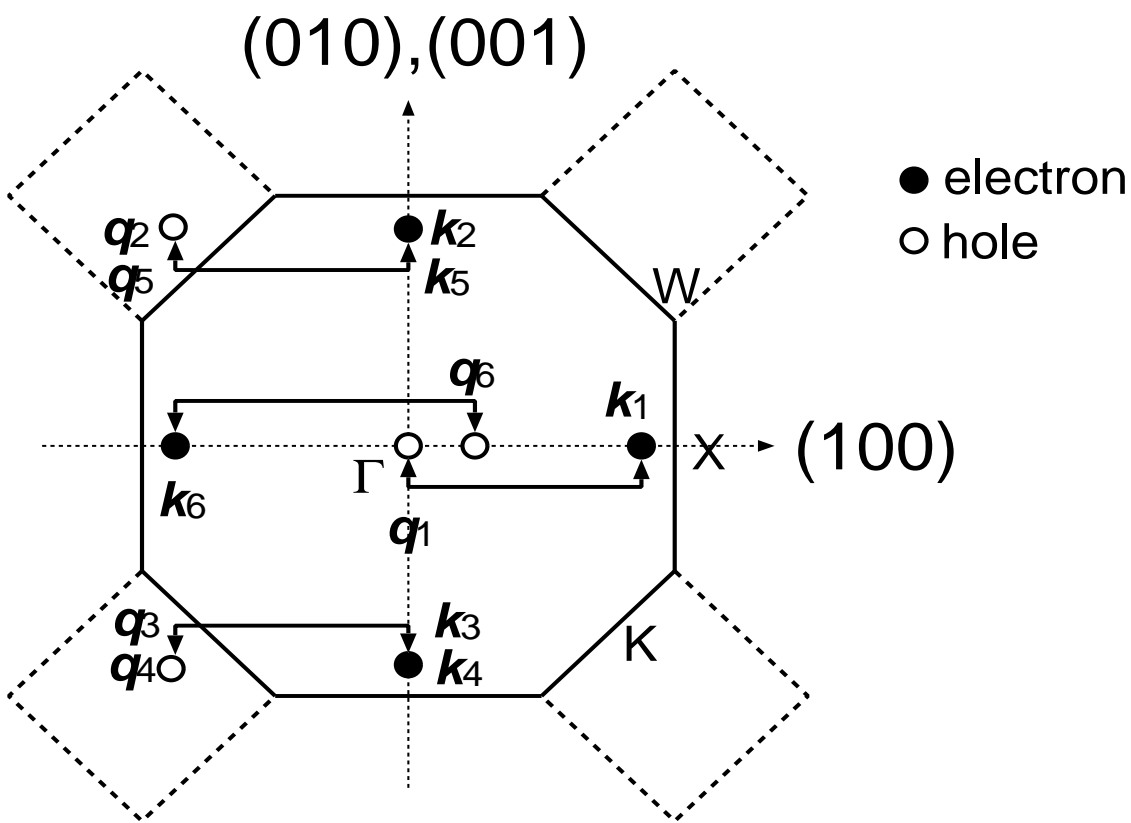}
\caption{Cross-section of the first Brillouin zone.
The crystal momenta of electrons and holes generated in final state are presented.}
\label{fig:BZ}
\end{figure}

\begin{figure}
\includegraphics[width=8cm,height=20cm]{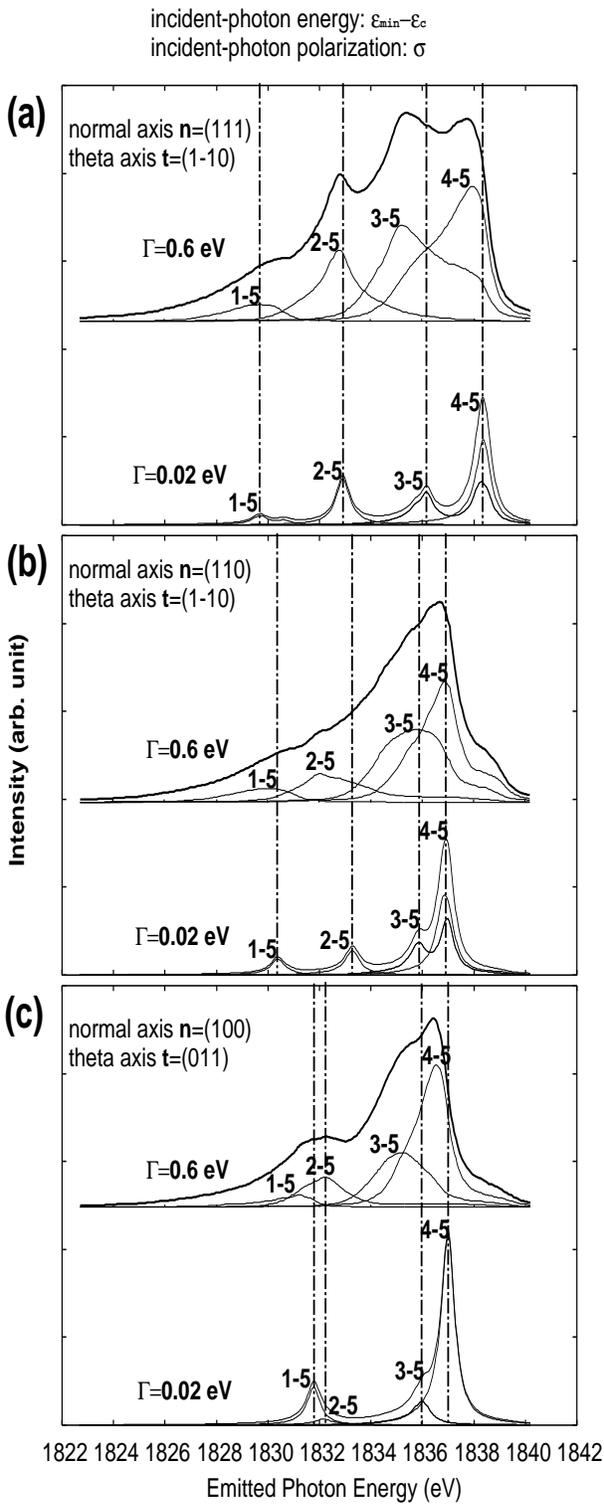}
\caption{RIXS spectra for realistic value of core-level
width ($\Gamma=0.6$ eV) in comparison with the RIXS spectra
for sufficiently small value of core-level width ($\Gamma=0.02$ eV).}
\label{fig:G0vsG0.6}
\end{figure}

\begin{figure}
\includegraphics[width=16cm,height=10cm]{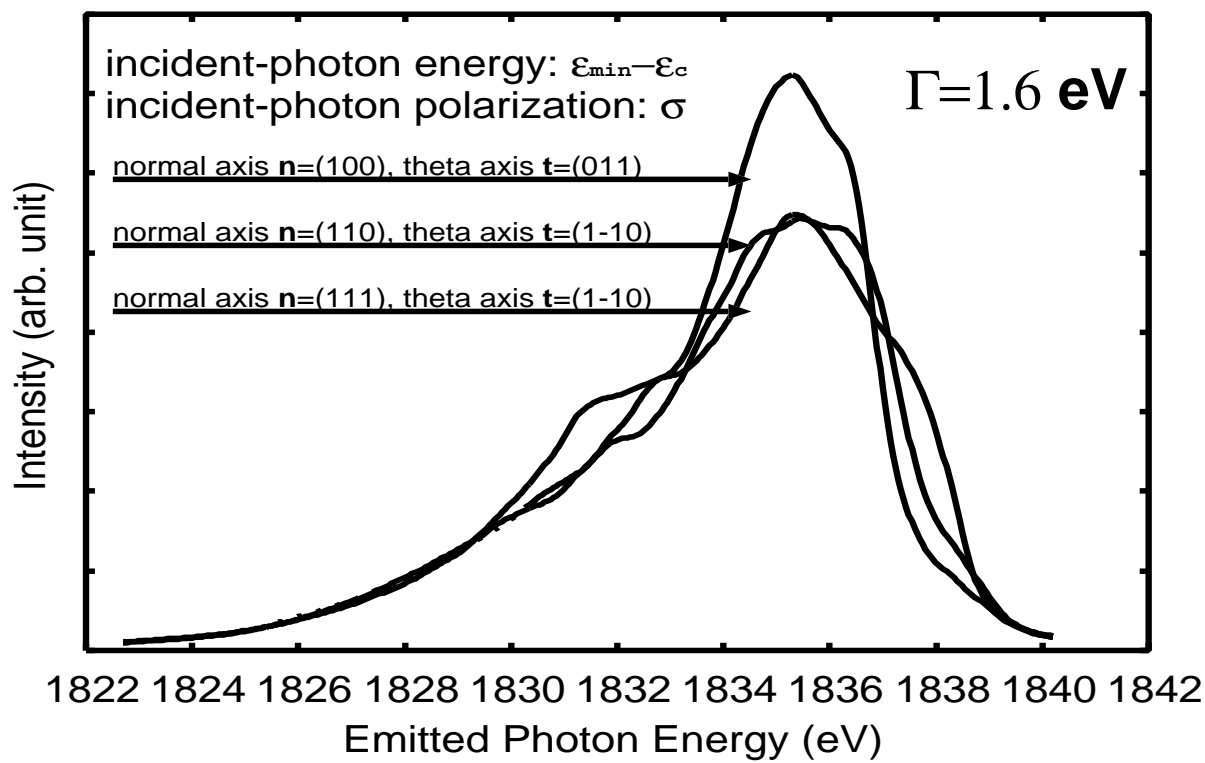}
\caption{RIXS spectra for large value of core-level
width ($\Gamma=1.6$ eV). The anisotropy almost disappears in the spectra.}
\label{fig:Gc}
\end{figure}
\clearpage
\begin{figure}
\includegraphics[width=17cm,height=10cm]{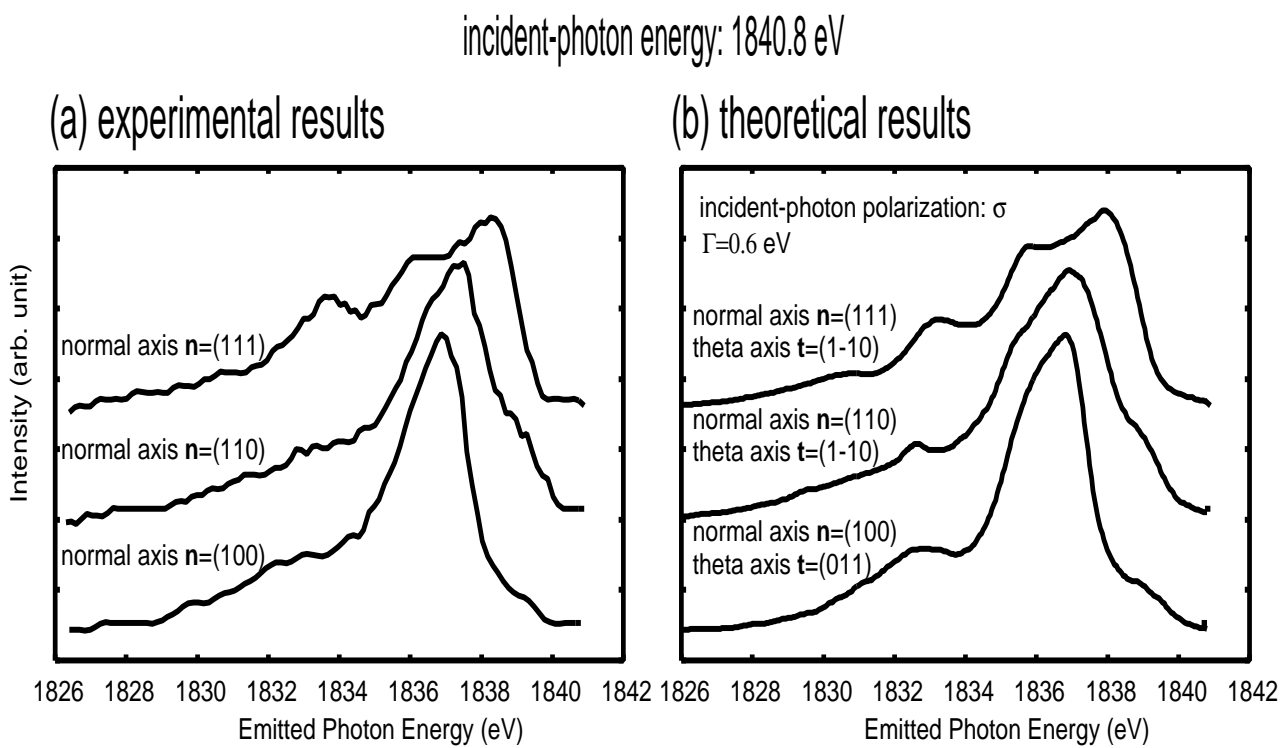}
\caption{RIXS spectra obtained by (a)experiment\cite{rf:MaExp}
 and (b)theory}
\label{fig:1840.8}
\end{figure}

\begin{figure}
\includegraphics[width=8cm,height=18cm]{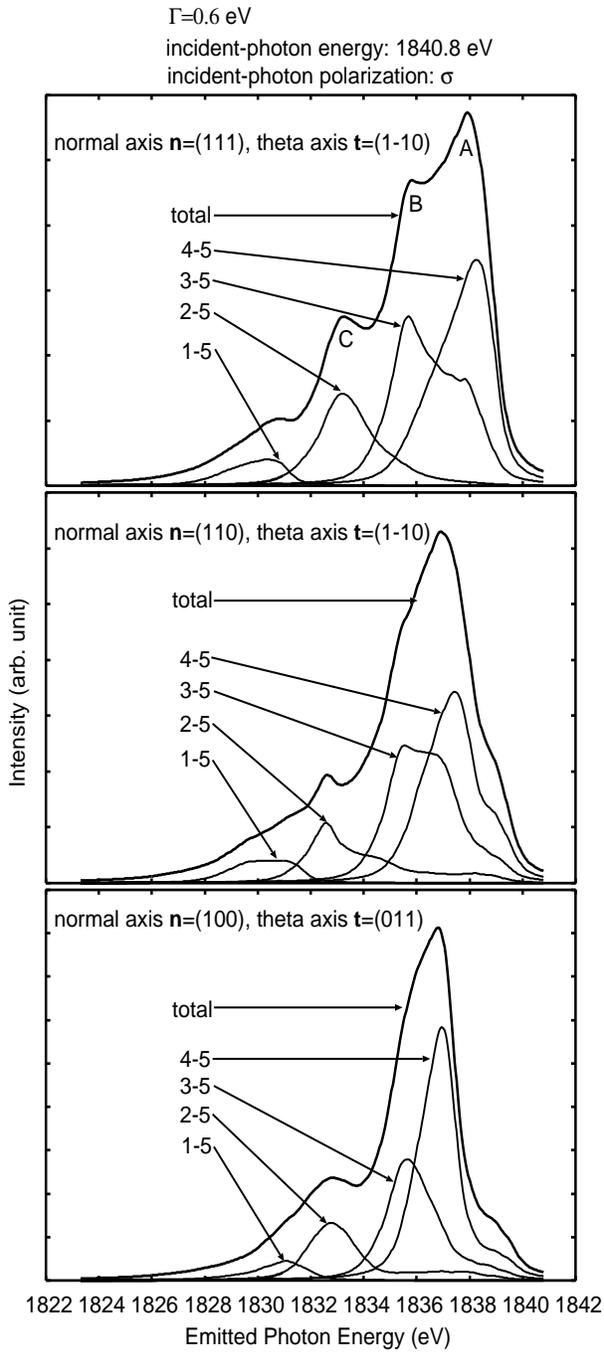}
\caption{The spectra for $\hbar\omega_{1}=$1840.8 eV,
${\bf n}$=(111), (110), and (100) are decomposed into each contribution
of band-to-band transitions specified by band indices.}
\label{fig:decomp}
\end{figure}

\clearpage
\begin{figure}
\includegraphics[width=15cm,height=11cm]{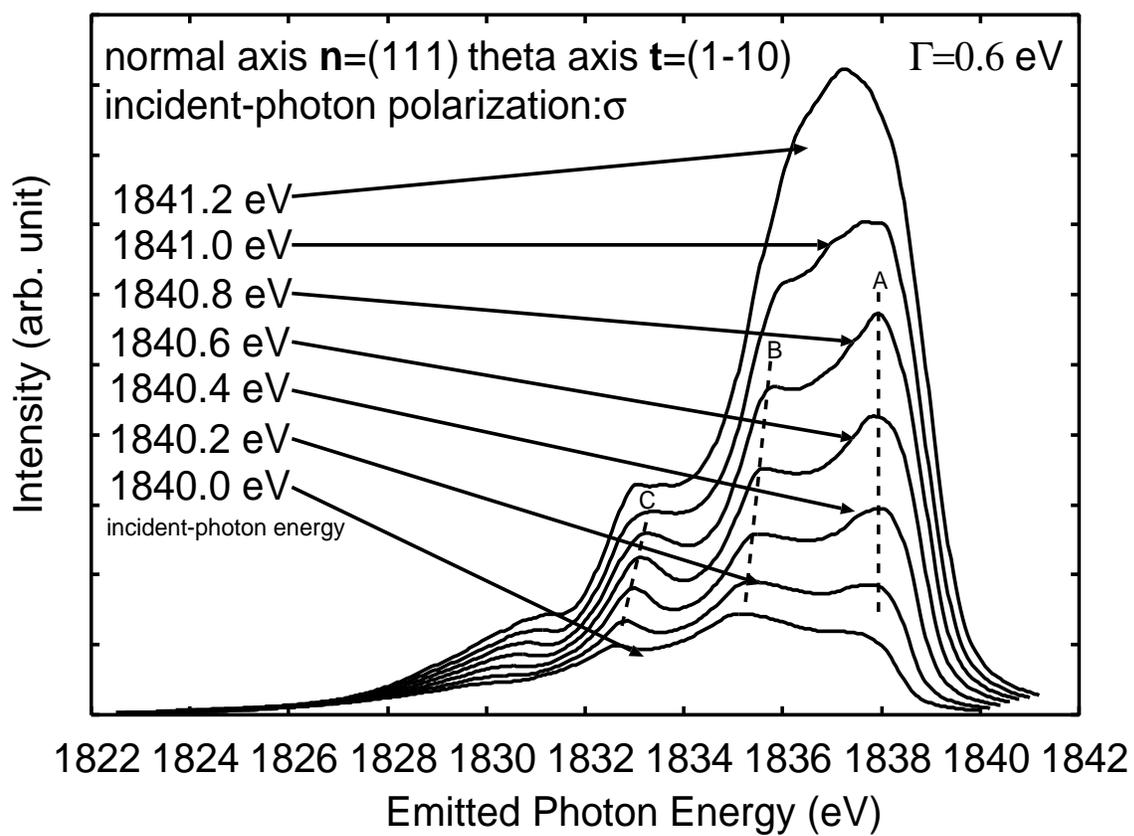}
\caption{Calculated RIXS spectra
for ${\bf n}$=(111) with varying incident-photon energy
between 1840.0 eV and 1841.2 eV.}
\label{fig:enedep111}
\end{figure}

\begin{figure}
\includegraphics[width=15cm,height=11cm]{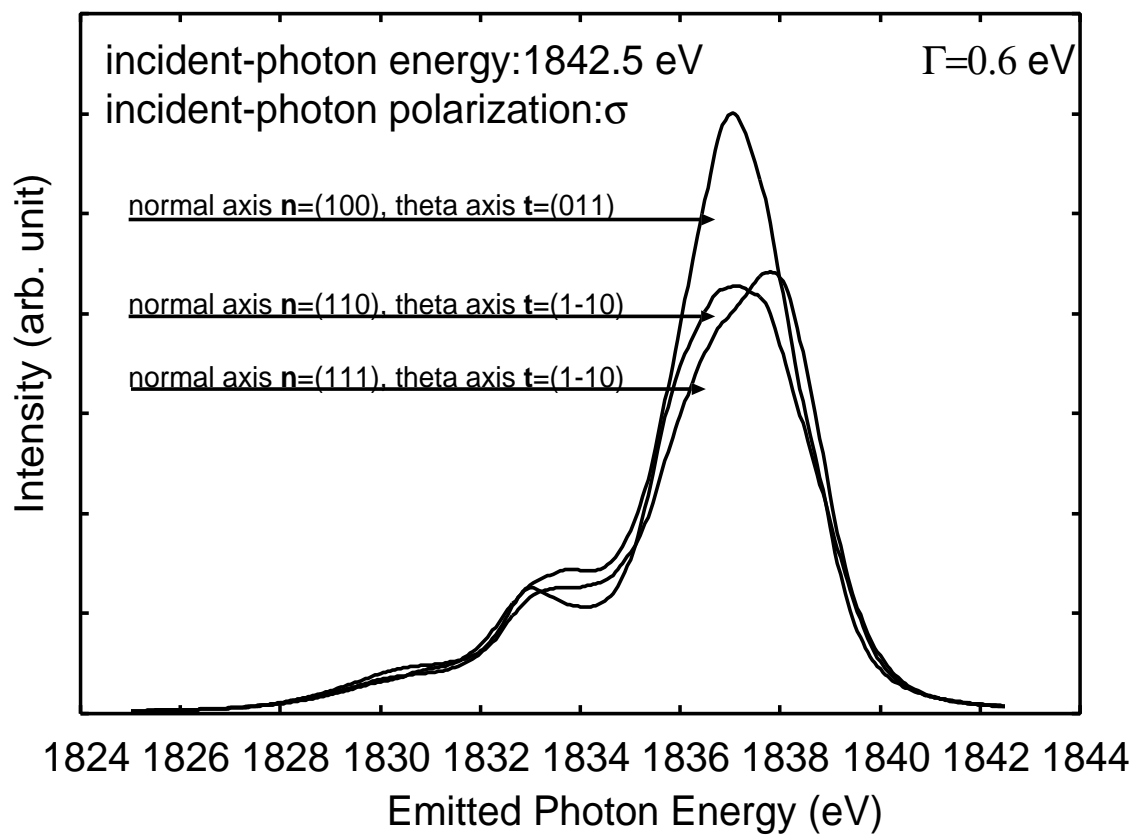}
\caption{Calculated RIXS spectra
for $\hbar\omega_{1}=1842.5$ eV with varying ${\bf n}$=(111), (110), and
 (100). }
\label{fig:enedephigh}
\end{figure}

\begin{figure}
\includegraphics[width=15cm,height=15cm]{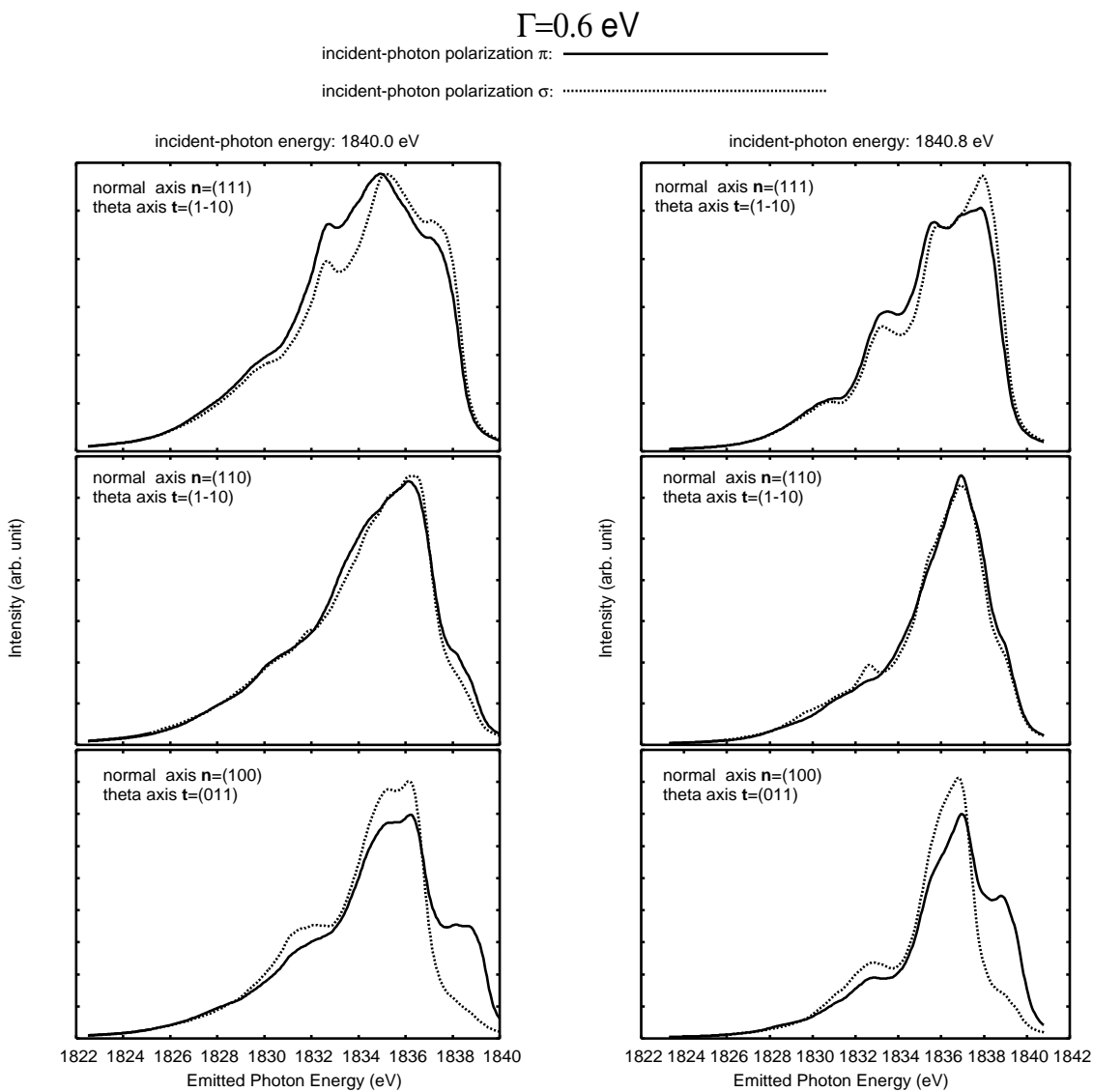}
\caption{Calculated RIXS spectra with varying incident-photon polarization.}
\label{fig:poldep}
\end{figure}

\begin{figure}
\includegraphics[width=15cm,height=20cm]{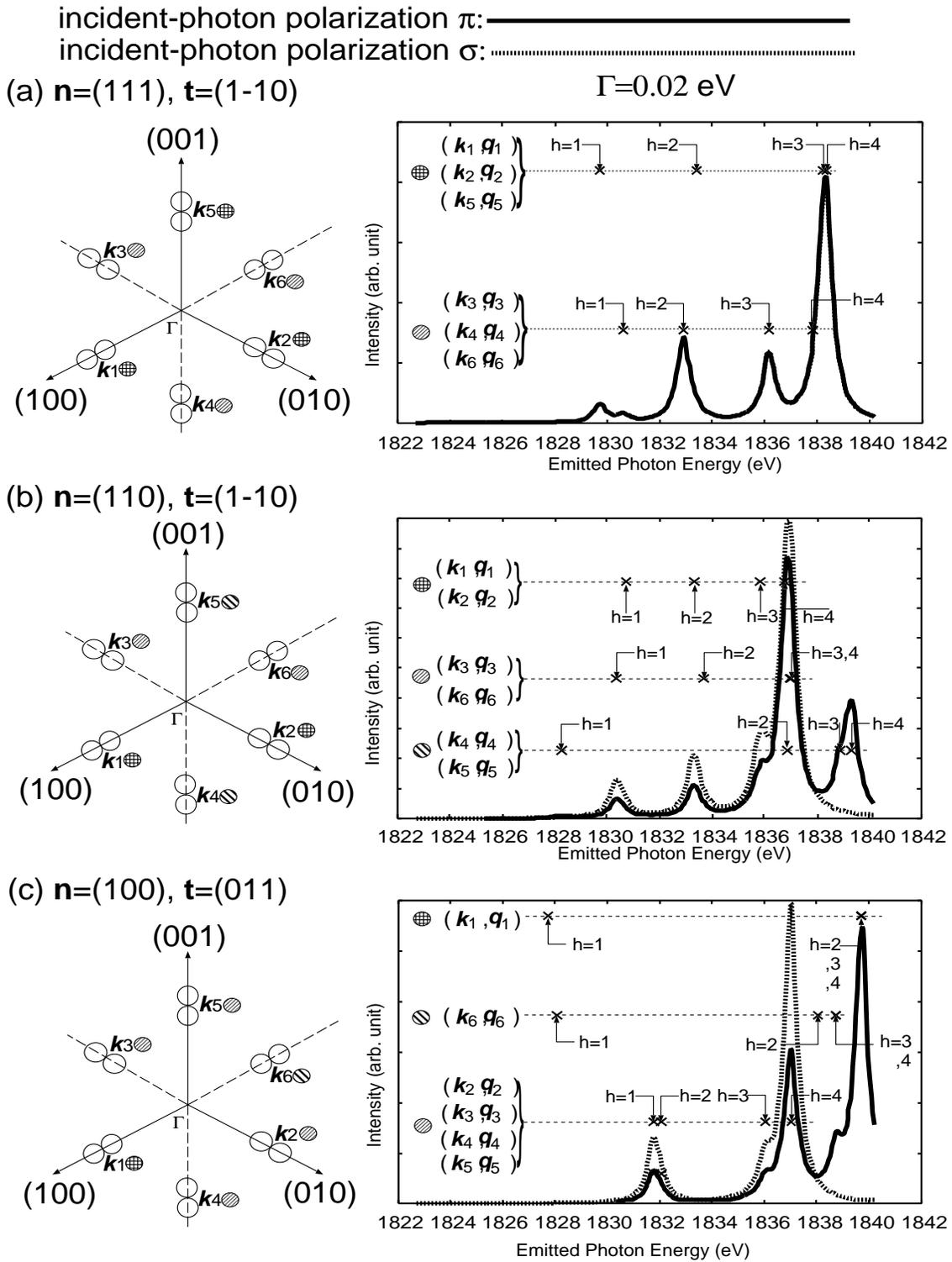}
\caption{Incident-photon polarization dependence of
RIXS spectra for $\hbar\omega_{1}=\epsilon_{\rm min}-\epsilon_{c}$
with sufficiently small value of $\Gamma$(=0.02 eV).
The positions of peaks expected to be made by
each electron-hole pairs are also presented.}
\label{fig:kstar}
\end{figure}


\end{document}